\documentclass[aps,prl,reprint,twocolumn,amsmath,amssymb,citeautoscript]{revtex4-1}
\usepackage{graphicx}
\usepackage{dcolumn}
\usepackage{bm}
\usepackage{latexsym,epsfig}
\usepackage{graphicx}
\usepackage{verbatim}
\usepackage{comment}
\usepackage{amsmath}
\usepackage{amssymb}
\usepackage{stmaryrd}
\usepackage{color}
\usepackage{epstopdf}
\usepackage{grffile}
\DeclareGraphicsExtensions{.eps}
\DeclareGraphicsExtensions{.eps}

\newcommand{\beq}{\begin{equation}}
\newcommand{\eeq}{\end{equation}}
\newcommand{\bea}{\begin{eqnarray}}
\newcommand{\eea}{\end{eqnarray}}
\newcommand{\ben}{\begin{eqnarray*}}
\newcommand{\een}{\end{eqnarray*}}
\newcommand{\bfig}{\begin{figure}}
\newcommand{\efig}{\end{figure}}

\begin{document}
\title{Topological inheritance in half-SSH Hubbard models}
\author{Suman Mondal$^1$, Sebastian Greschner$^2$, Luis Santos$^3$ and Tapan Mishra$^1$}
\affiliation{$^1$Department of Physics, Indian Institute of Technology, Guwahati-781039, India}
\affiliation{$^2$ Department of Quantum Matter Physics, University of Geneva, 1211 Geneva, Switzerland}
\affiliation{$^3$Institute for Theoretical Physics, Leibniz University of Hannover, Hannover, Germany}
\date{\today}

\begin{abstract}
The interplay between interparticle interactions and topological features may result in unusual phenomena. Interestingly, interactions may induce 
topological features in an originally trivial system, as we illustrate for the case of  a one-dimensional two-component Hubbard model in which one component is subjected to 
Su-Schrieffer-Heeger(SSH) dimerization, whereas the other one is not. We show that due to inter-component interactions 
the topological properties of one component are induced in the originally trivial one. Although for large interactions topological inheritance may be readily explained 
by on-site pairing, we show that the threshold for full inheritance occurs at weak interactions, for which the components are not yet paired. 
We illustrate this inheritance by discussing both bulk and edge properties, as well as dynamical observables as mean chiral displacement and charge pumping.
\end{abstract}


\maketitle



Symmetry protected topological~(SPT) phase transitions are a class of phase transitions 
which do not come under the well-known Landau-Ginzburg paradigm associated with symmetry breaking. After the seminal observation of the fractional quantum Hall effect in 
condensed-matter systems~\cite{Klitzing1980,Tsui1982,Thouless_hall,Laughlin1983,StormerTsui1983,Stormer_Rev1999}, 
research on SPT phases in disparate systems ranging from exotic materials, ultracold neutral atoms, trapped ions, or photonic systems has 
earned enormous attention in recent years~\cite{Rachel_TopRev2018,Zhang_Rev2011,KaneColloquium2010,CooperRev2019,Nevado2017,Mezzacapo2013,LeHur2016,TopPhotRevOzawa2019}. 

One of the simplest models exhibiting an SPT phase transition is the Su-Schrieffer-Heeger~(SSH) model,  
which was first discussed in the context of solitons in polyacetylene~\cite{ssh}. The SSH model is a two-band one-dimensional 
tight-binding model with dimerized hoppings, which exhibits a topologically trivial to non-trivial transition through a gapless point. The non-trivial phase possesses
zero-energy edge modes and a non-zero quantized Zak phase~\cite{Zak1989}. 
The non-interacting SSH model has been extensively analysed as a paradigmatic model to 
understand topological phenomena in various systems~\cite{Bello2019,Engelhardt2017,Downing2017,Henning2013,Chaunsali2017,Kane2014,Grusdt2013}. It 
has been recently implemented in quantum gas and photonic experiments~\cite{Atala2013,Meier2016,TopPhotRevOzawa2019,Malkova2009,St-Jean2017} and, in particular, 
Thouless topological charge pumping~\cite{Thouless1983} has been observed~\cite{Lohse2015,Takahashi2016pumping,Schweizer2016,Lohse2018,Goldman2013,Silberberg2013,Silberberg2015}.
Interactions lead to exciting new physics due to the competing interplay of correlation effects, particle statistics, and lattice topology, as recently discussed theoretically 
for both Bose-~\cite{Grusdt2013,DiLiberto2016,Suman2019} and Fermi-Hubbard models~\cite{Yoshida2018,Barbiero2018,Wang2015,Ye2016,Dalmonte2020,Li2013}.

In this paper, we show how, due to interactions, a topological system may induce topological features in a non-topological one. In particular, 
we consider a two-component system, in which one of the components experiences dimerized hopping, and hence an SSH model,  
whereas the other presents non-dimerized hopping, and would be hence topologically trivial if considered alone.
We show that a finite interaction~(attractive or repulsive) coupling the two species maps topological properties into the a-priori non-topological component, resulting in the formation of 
strongly-correlated edge pairs.



\begin{figure}[t]
\begin{center}
\includegraphics[width=0.7\columnwidth]{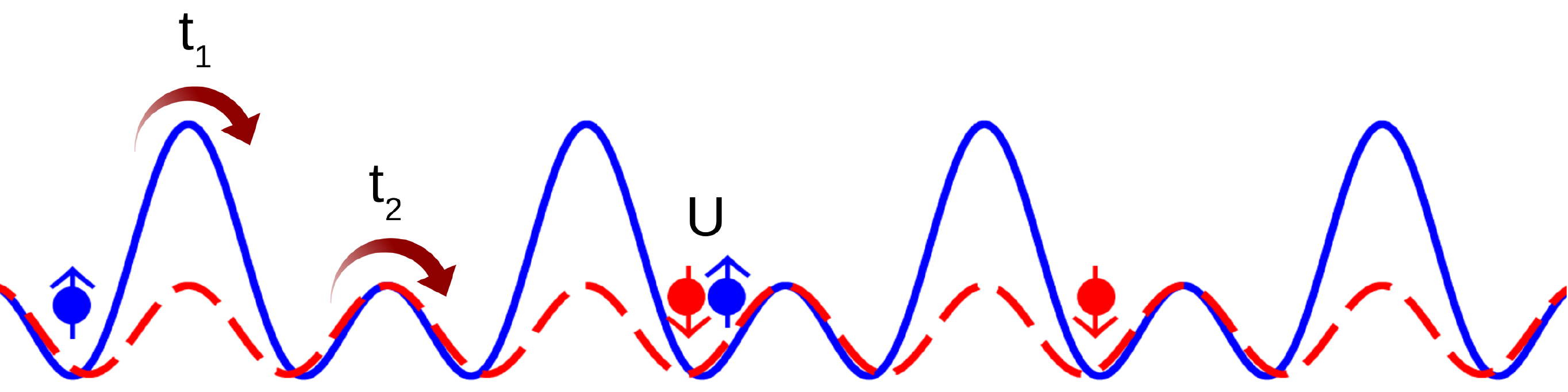}
 \end{center}
 \vspace*{-0.5cm}
\caption{(Color online) Two different species $\{\uparrow, \downarrow\}$ in the Model~\eqref{eq:hsshh}. Whereas $\uparrow$ experiences SSH hopping dimerization with $t_1\neq t$, 
$\downarrow$ is in a homogeneous lattice with uniform hopping amplitude $t$. The inter-component interaction is denoted as $U$.}
\label{fig:lattice}
\end{figure}

\paragraph{Model.--} We consider two Bose components $\{\uparrow,\downarrow\}$, such that $\uparrow$ experiences SSH dimerization, whereas $\downarrow$ is not dimerized. 
The interacting many-body system is described by the model:
\begin{align}
\mathcal{H} = &- t_1\!\! \sum_{i\in odd} \!\!(c_{i\uparrow}^{\dagger}c_{i+1\uparrow}^{\phantom \dagger} + \text{H.c.})
- t_2\!\!\!\sum_{i\in even} \!\! (c_{i\uparrow}^{\dagger}c_{i+1\uparrow}^{\phantom \dagger}+\text{H.c.})\nonumber\\
 &- t\sum_{i}(c_{i\downarrow}^{\dagger}c_{i+1\downarrow}^{\phantom \dagger}+\text{H.c.})
  + U\sum_{i}  n_{i\uparrow}n_{i\downarrow}
\label{eq:hsshh}
\end{align}
where $c_{i\sigma}^{\dagger}$ and $c_{i\sigma}$ are the creation and annihilation operators  for $\sigma=\uparrow,\downarrow$ at site $i$, 
$n_{i\sigma}=c_{i\sigma}^{\dagger}c_{i\sigma}^{\phantom \dagger}$ are the number operators, 
$t_1$ and $t_2$ are the tunneling rates of $\uparrow$ from odd and even sites, respectively, $t$ is the hopping rate of $\downarrow$, and 
$U$ characterizes the inter-component interaction. We consider the hard-core constraint, $n_{\uparrow,\downarrow}\leq 1$. 
Note that model~\eqref{eq:hsshh} can be mapped to a Fermi mixture, which presents the same 
spectrum and diagonal correlations.
 
In absence of interactions~($U=0$) model~\eqref{eq:hsshh} reduces to two uncoupled models, an SSH model for $\uparrow$, and a trivial Hubbard model for $\downarrow$. The quantized 
Zak phase of $\uparrow$ is zero for $t_1>t_2$~(trivial phase) and $\pi$ for $t_1 < t_2$~(topological phase). While the bulk remains gapped for both phases, only the latter  possesses zero energy edge modes. 
In our density matrix renormalization group~(DMRG) calculations below, we consider half-filling, in which the number of particles in both components $N_{\uparrow,\downarrow}=L/2$, for a lattice with $L$ sites.
We set $t_2=t=1$ as energy unit, and fix $t_1=0.2$, within the non-trivial regime for $\uparrow$. 



\begin{figure}[t]
\begin{center}
\includegraphics[width=1\columnwidth]{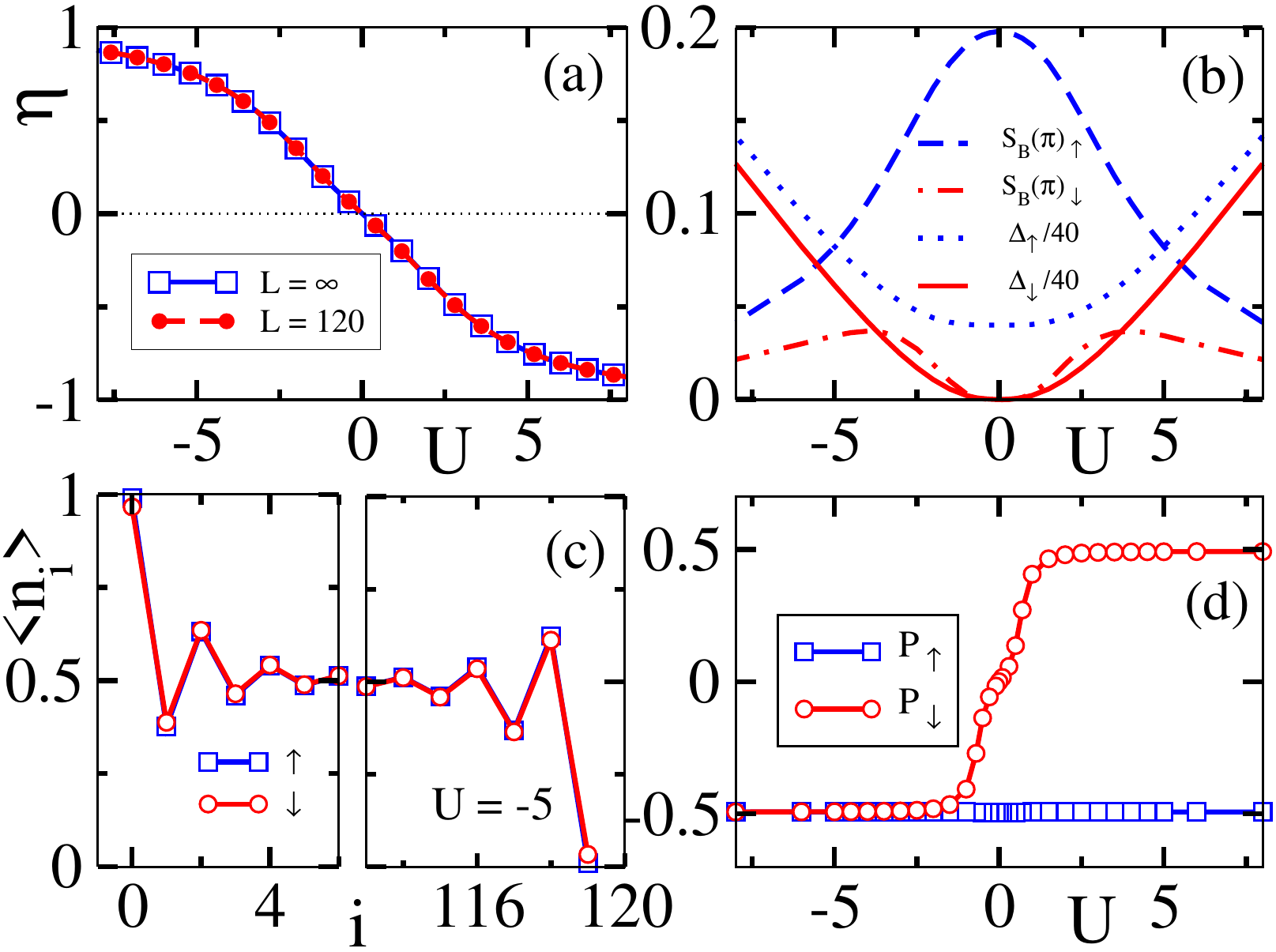}
 \end{center}
 \vspace*{-0.5cm}
\caption{(Color online) (a) Pairing $\eta$ as a function of $U$; (b) Bulk gaps $\Delta_\sigma$~(blue dotted and red continuous), and dimer structure factors $S_{D\sigma}(\pi)$ (blue dashed and red dot-dashed) for $\uparrow$ and $\downarrow$ components, respectively, as a function of $U$. The observables are extrapolated to $L\rightarrow\infty$ using our DMRG results for system sizes of length $L = 40,~60,~80,~100$ and $120$; (c) Edge states for both component for $U=-5$ for a system with $L=120$; 
(d) Polarization $P_\sigma$ for $\uparrow$~(blue squares) and $\downarrow$~(red circles) as a function of $U$, obtained for $L=120$.}
\label{fig:strdimer}
\end{figure}

\paragraph{Pairing.--} For a sufficiently strong $U<0$~($>0$) a particle of one component pairs on-site with a particle~(hole) of the other. Pairing is best monitored by $\eta=\frac{4}{L}\sum_i \left ( \langle n_{i\uparrow}n_{i\downarrow}\rangle-\frac{1}{4} \right )$, see Fig.~\ref{fig:strdimer}(a). For large-enough 
 $|U|> 8$, $|\eta|\simeq 1$ indicating strongly localized on-site pairing. 
For strong particle-particle pairing~(particle-hole pairing is treated analogously), 
the system is described by an effective SSH model for on-site hard-core pairs:
\begin{equation}
\mathcal{H} =\frac{2t_1 t}{U} \sum_{i\in odd} \!\! P_{i}^{\dagger}P_{i+1}^{\phantom \dagger} + 
\frac{2t_2 t}{U} \sum_{i\in even} \!\! P_{i}^{\dagger}P_{i+1}^{\phantom \dagger} + \text{H.c.},
\label{eq:Heff}
\end{equation}
with $P_i=c_{i\uparrow}c_{i\downarrow}$. Model~\eqref{eq:Heff} is topological if the SSH model for $\uparrow$ is topological. 
Hence, the pairs (and with them the second component) inherit the topology of the first component.
Interestingly, as discussed below, much weaker pairing $\eta\ll 1$, and hence very moderate $|U|$, already suffices to induce 
full {\em topological inheritance}.

\paragraph{Induced bulk properties.--}  In absence of interactions, the $\downarrow$ component is a gapless superfluid, whereas due to hopping dimerization 
the bulk of the $\uparrow$ component is in a gapped dimer phase with a finite dimer structure factor,
$S_{D\sigma}(k)=\frac{1}{L^2}\sum_{i,j}e^{ikr}\langle D_{i\sigma}D_{j\sigma}\rangle$, with 
$D_{i\sigma}= c_{i\sigma}^{\dagger}c_{i+1\sigma} + \mathrm{H. c.}$
The dotted blue~(solid red) curve in Fig.~\ref{fig:strdimer}(b) depicts the charge gap $\Delta_{\sigma}=E(L, N_\sigma+1)+E(L, N_\sigma-1) -2E(L, N_\sigma)$ and dashed blue~(dot-dashed red) curve depicts the dimer structure factor for 
both components. Strong pairing asymptotically results for large $|U|$ in an exact replication of the bulk properties of the $\uparrow$ component on the $\downarrow$ component. Note, however, that any $U\neq 0$~(either repulsive or attractive) results in a finite bulk gap and dimer order in the spin-$\downarrow$ component. 
Any finite interaction drives the bulk of the $\downarrow$ component into a gapped dimer phase.



\paragraph{Induced edge states--} For $t_1<t_2$ the $\uparrow$ component is non-trivial, possessing doubly-degenerate edge modes. 
Sufficiently attractive~(repulsive) interactions induced correlated~(anti-correlated) edge states in $\downarrow$, see Fig.~\ref{fig:strdimer}~(c).
The inheritance of the edge states by the $\downarrow$ component is best monitored by the polarization 
$P_\sigma = \frac{1}{L} \sum_{i=0}^{L} \langle \psi | (i-L/2) n_{i\sigma} | \psi \rangle$ for  the 
ground state $|\psi\rangle$, which we depict in Fig.~\ref{fig:strdimer}(d) for $L=120$ for different values of $U$.
Note that $P_\uparrow=-1/2$ due to the topological character of the $\uparrow$ component.
The $\downarrow$ component shows maximally polarized edges, $|P_\downarrow|\simeq 1/2$, already for $|U|\sim 1$.
Hence, the $\downarrow$ component fully inherits the topological edge modes for values of $|U|$ well below those needed for strong pairing. 


 
\paragraph{Mean chiral displacement.--}  Topological inheritance may be easily probed experimentally by monitoring the dynamical evolution after a quench.
The mean-chiral displacement~(MCD), recently utilized in the context of quantum random walk on a graph, can be utilized to 
measure the topological winding number in photonic and ultracold atomic systems ~\cite{Meier2018,LewensteinMCD,Xie2019,Xie2020}. 
The MCD is defined as $C_\sigma(\tilde{t})=2\langle \Psi(\tilde{t}) | \Gamma_\sigma m_\sigma | \Psi(\tilde{t}) \rangle$,  where $\Gamma_\sigma$ and 
$m_\sigma$ are the chiral and unit cell operators, respectively, and  $|\Psi(\tilde{t})\rangle$ is the time-evolved state. 
The MCD displays an oscillatory behavior, but after a sufficiently large time its time averaging converges
to the winding number $\omega$~\cite{supmat}.

\begin{figure}[b]
\begin{center}
\includegraphics[width=1\columnwidth]{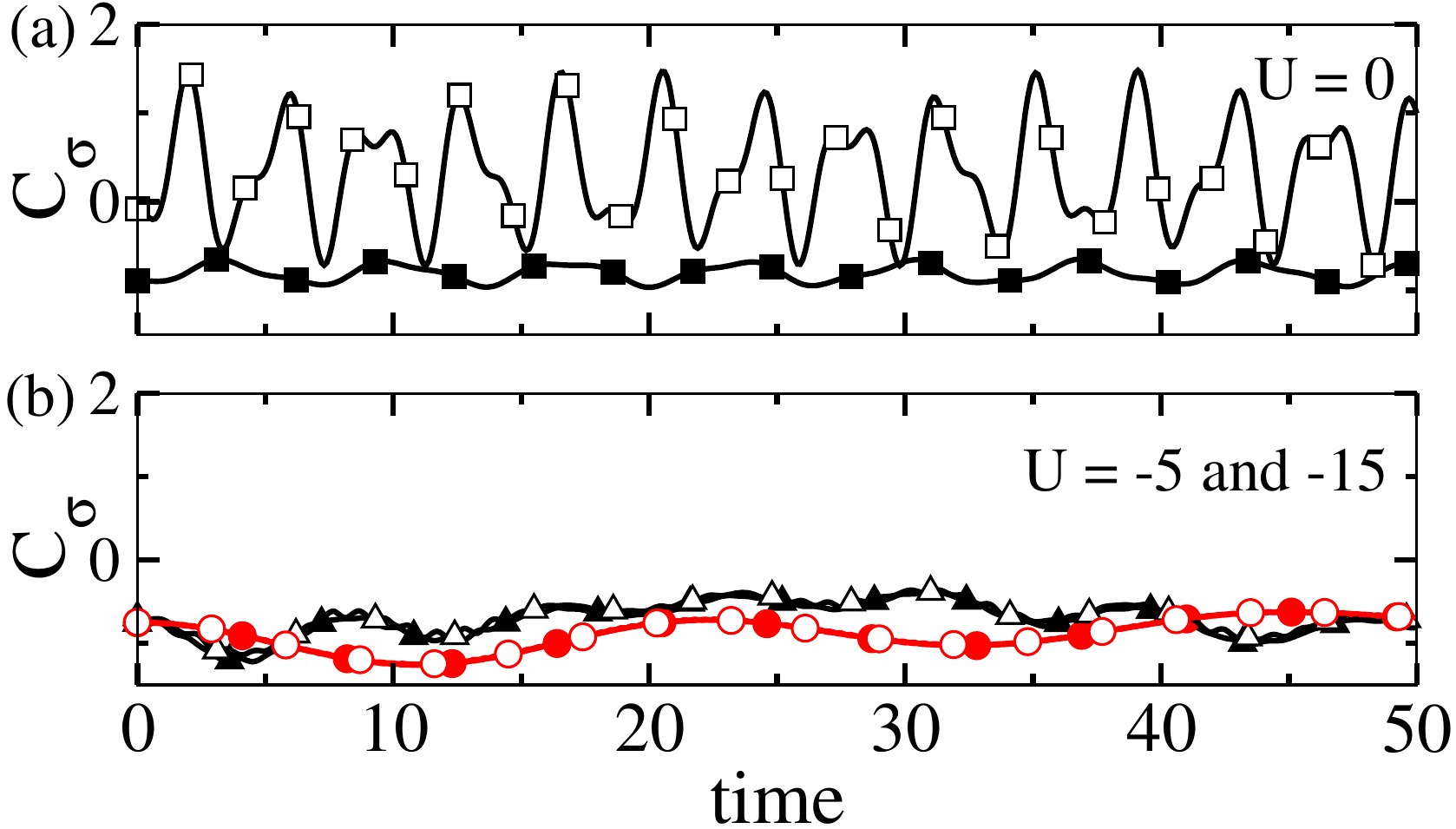}
 \end{center}
 \vspace*{-0.5cm}
\caption{(Color online) (a) MCD for a system with $L=6$ for $U=0$ for $\uparrow$~(filled symbols) and $\downarrow$~(hollow symbols); (b) 
Same for $U=-15$~(red circles) and $U=-5$~(black triangles).
Here solid and empty symbols are corresponding to $\uparrow$ and $\downarrow$ respectively. For $|U|=5$($15$) the MCD evolution curves are marked by symbol triangle(circle). 
In all cases we employ as initial state $|\Psi(0)\rangle = c_{0\uparrow} c_{0\downarrow}|\Psi_{GS}\rangle$.}
\label{fig:mcd}
\end{figure}

We analyze the MCD by 
considering an initial state $|\Psi(0)\rangle = O|\Psi_{GS}\rangle$, with $|\Psi_{GS}\rangle$ the ground-state, and $O=c_{0\uparrow}c_{0\downarrow}$~($O=c_{0,\uparrow}^\dag c_{0,\downarrow}$), i.e. 
pair annihilation~(spin flip) at the central site, for $U<0$~($>0$).
Quantum walks from these initial states provide insights about the charge and spin winding numbers, respectively. 
Figures~\ref{fig:mcd} show our results of $C_\sigma(\tilde{t})$, evaluated with the pair annihilation, for a system of size $L=6$ with open boundary conditions. 
For $U_{\uparrow,\downarrow}=0$~(Fig.~\ref{fig:mcd}(a)), $C_\uparrow(\tilde{t})$~(filled squares) oscillates around the winding number $-1$, as expected from 
the topological character of the $\uparrow$ component. 
In contrast, $C_\downarrow(\tilde{t})$ oscillates around zero~(open squares), showing no topological behavior. 
When increasing the inter-component interaction, the MCD of both components becomes identical oscillating around the winding number~(Fig.~\ref{fig:mcd}(b))
revealing the inheritance by the $\downarrow$ component of the topological properties of the $\uparrow$ component.
Note that the deviations from the true winding numbers can be attributed to the finite size and interaction effects~\cite{Matteo2020}.



\begin{figure}[t]
\begin{center}
\includegraphics[width=1\columnwidth]{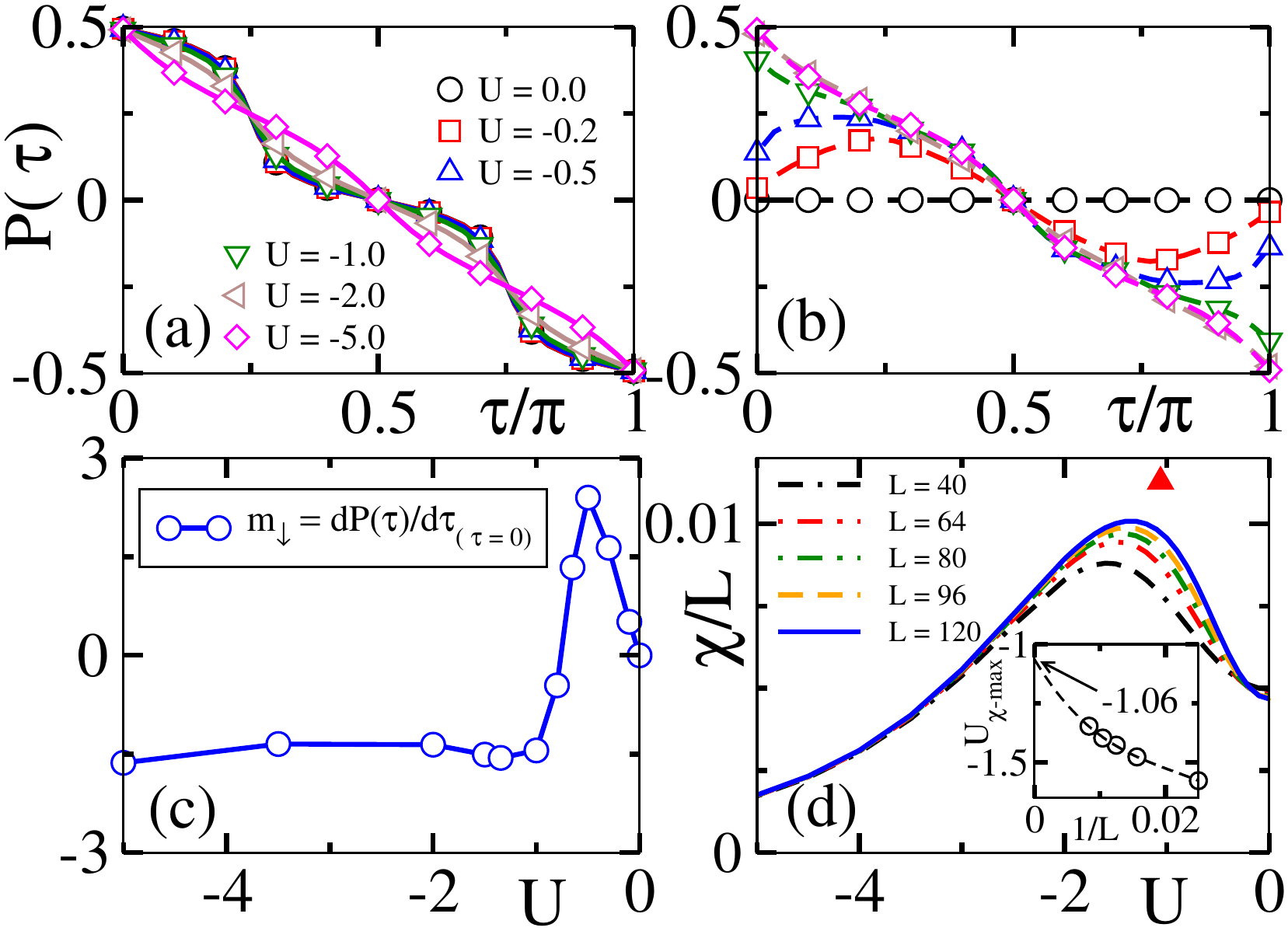}
 \end{center}
 \vspace*{-0.5cm}
\caption{(Color online) 
Polarizations $P_\uparrow(\tau)$~(a) and $P_\uparrow(\tau)$~(b) for different $U$ values and $L=120$;
(c) Initial gradient $m_\downarrow=\frac{dP\downarrow}{d\tau}(\tau=0)$ as a function of $U<0$; (d) Fidelity susceptibility $\chi$ as a function of $U$ for different $L$. The inset shows the extrapolation of the 
value $U$ at which $\chi/L$ has a maximum. This value marks the onset of full topological inheritance. Here, the filled red triangle corresponds to the extrapolated position of the peak.
}
\label{fig:pump}
\end{figure}

\paragraph{Thouless charge pumping.--}
Alternatively, topological inheritance may be dynamically probed by investigating Thouless charge pumping, i.e. the transport of quantized charge as a result of an adiabatic periodic modulation of the system parameters. 
We consider that only the $\uparrow$ component is driven following a Rice-Mele~(RM) model~\cite{Rice1982}  
\begin{eqnarray}
{\cal H}_{\uparrow} &=& -\sum_i \left [ (t - (-1)^i \delta t \cos(2 \tau)) c_{i,\uparrow}^\dagger c_{i+1, \uparrow} + {\rm H.c.} \right ] \nonumber\\
&+& \frac{\delta \Delta}{2} \sin(2 \tau) \sum_i (-1)^i n_{i,\uparrow}\,.
\label{eq:HRM}
\end{eqnarray}
where $\tau$ is a cyclic parameter used for the pumping protocol. 
The RM model reduces for $\tau=\pi/4$ and $3\pi/4$ to the SSH model~\cite{Hayward2018,Suman2019}. The pumping is best evaluated by monitoring 
the polarizations $P_\sigma(\tau) = \frac{1}{L} \sum_{i=0}^{L} \langle \psi(\tau) | (i-L/2) n_{i\sigma} | \psi(\tau) \rangle$. The total charge pumped during the cycle can be obtained by computing $Q_\sigma = \int_0^1 d\tau \partial_\tau P_\sigma(\tau)$.  As expected, $P_\uparrow(\tau)$~(Fig.~\ref{fig:pump}(a)) shows the robust pumping of one particle, indicating the existence of edge states. 
While no pumping occurs in $\downarrow$ for $|U|=0$, increasing $|U|$ leads to a finite 
pumping~(Fig.~\ref{fig:pump}(b)), despite the fact that only $\uparrow$ is externally modulated. For $|U|\gtrsim 1$ the pumping of a full $\downarrow$ particle marks the 
complete topological inheritance.



\paragraph{Inheritance threshold.--} The previous results show that the $\downarrow$ component fully inherits the topological properties of the $\uparrow$ component~(edge states, winding number, MCD, 
full-particle Thouless pumping) for inter-particle interactions beyond a given threshold. Such an inheritance threshold is not only revealed by the edge polarization~(Fig.~\ref{fig:strdimer}~(d))
and the change of character of the charge pumping~(Figs.~\ref{fig:pump}~(b) and~(c)), but also by the analysis of the fidelity susceptibility, 
$\chi= \lim_{(U-U')\to 0}\frac{-2\ln |\langle\psi_0(U)|\psi_0(U'\rangle|}{(U-U')^2}$. 
As shown in Fig.~\ref{fig:pump}~(d), $\chi/L$ shows a clear maximum, that marks the inheritance threshold. Such a threshold 
approaches asymptotically $|U|\simeq 1.06$ for growing $L$. At the threshold, the pairing correlation $\eta\simeq 0.2$~(Fig.~\ref{fig:strdimer}~(a)), and hence, remarkably, the on-set 
of full topological inheritance occurs when the components are not yet paired.

\paragraph{Conclusions.--} Due to interactions, a topological system may induce topological features in a non-topological one, as we have illustrated for a Hubbard model 
in which a component experiences SSH dimerization and the other not. Although, for strong interactions topological inheritance may be readily understood from the formation of 
on-site localized inter-component pairs which experience an effective SSH model, we have shown that, interestingly, the threshold for full topological inheritance occurs for much weaker 
interactions, for which the two components are not yet paired.

\begin{figure}[t]
\includegraphics[height=4.9cm]{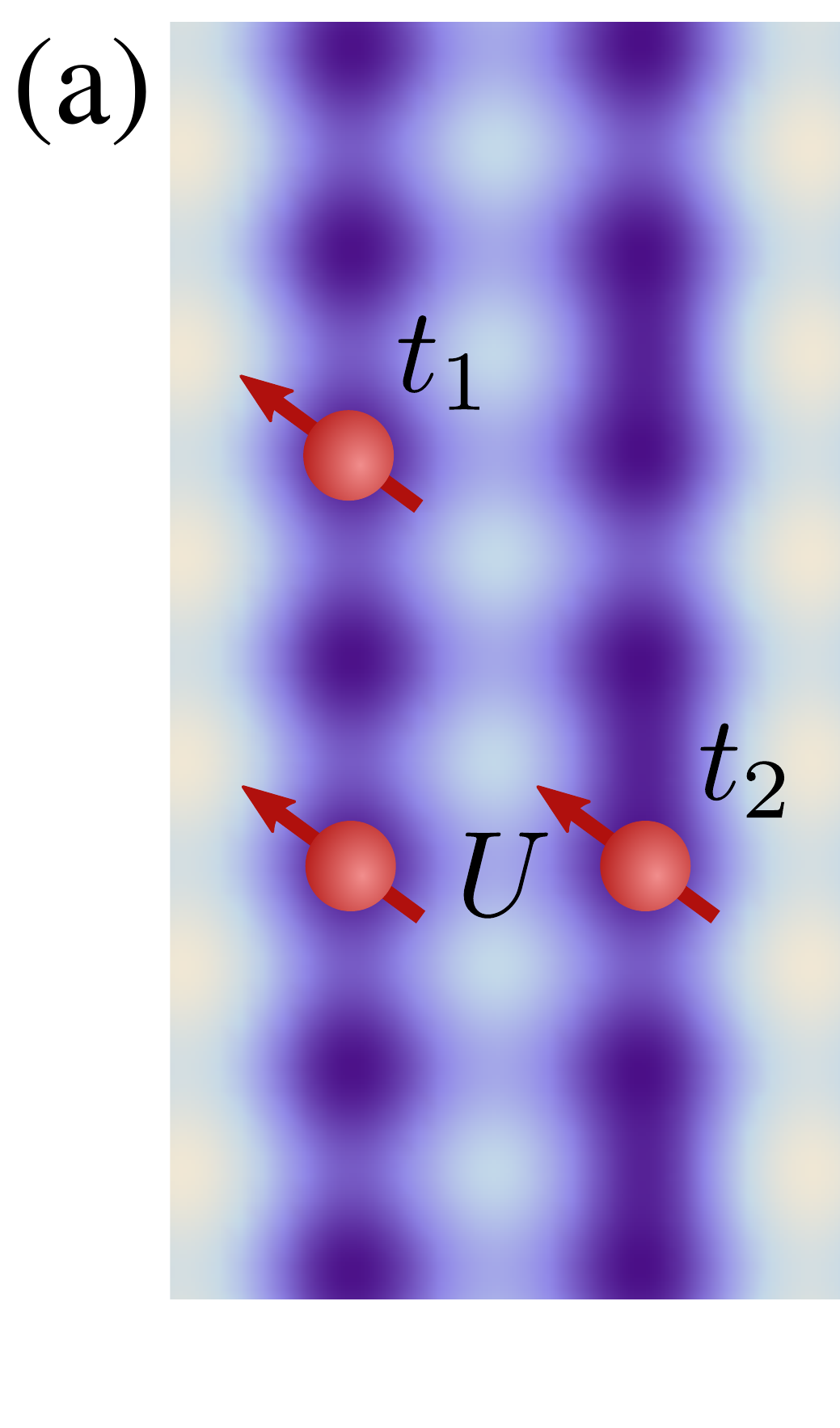}
\hspace{0.2cm}
\includegraphics[height=4.9cm]{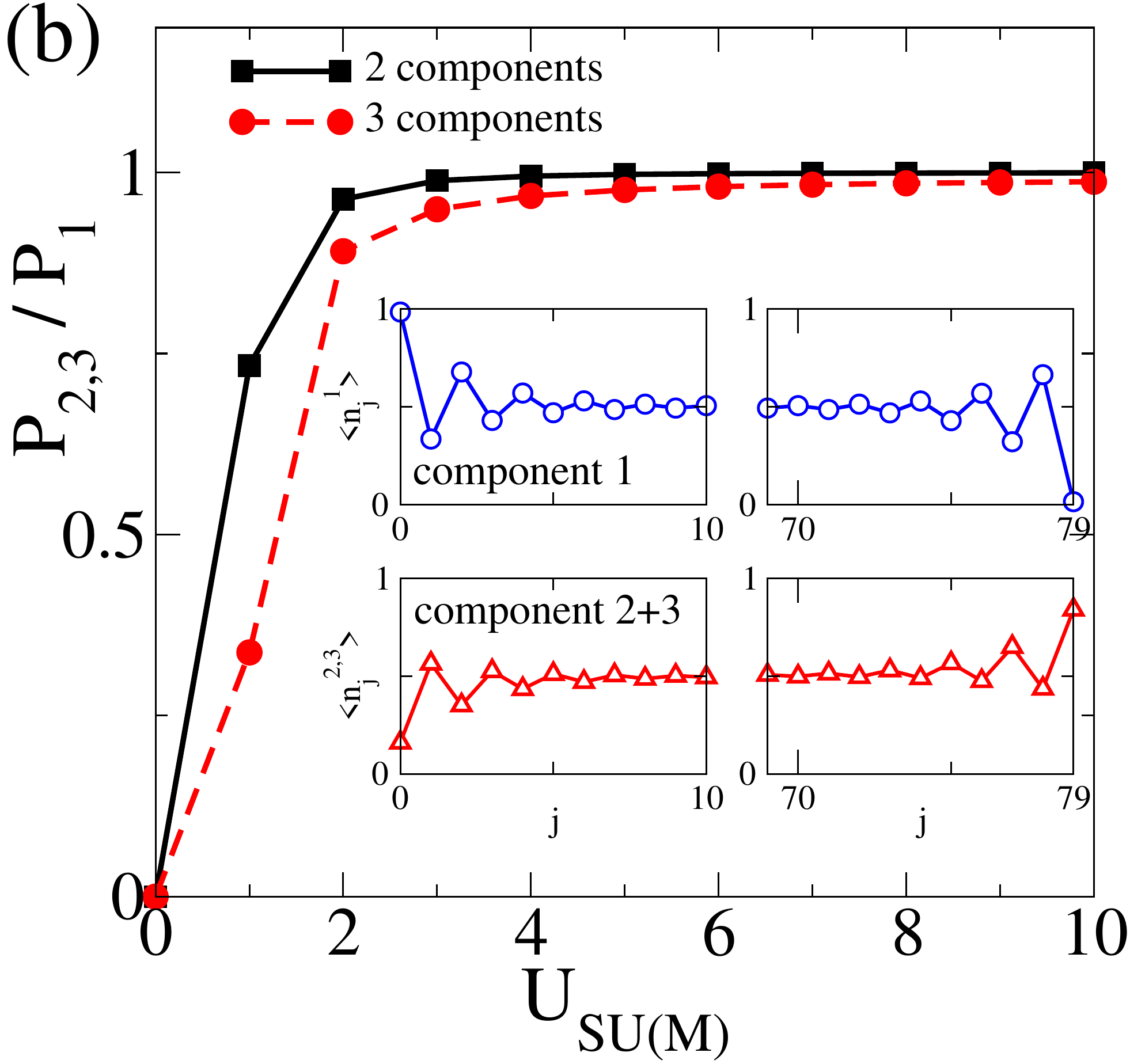}
\caption{(Color online) (a) Experimental scheme to realize Model~\eqref{eq:hsshh} using two-leg
optical ladder systems of hardcore dipolar bosons using dipolar particles~(see text); 
(b) Induced polarization for a three-species system in a
state-dependent super-lattice in the presence of on-site all-to-all
(SU(3)-symmetric) interactions. The inset shows the replication of the edge
state to the second and third species for $U_{SU(3)}=6$.}
\label{fig:scheme}
\end{figure}

A possible experimental realization of Model~\eqref{eq:hsshh} may be achieved by employing two neighboring 1D lattices of 
hardcore dipolar bosons, which act like the two components $\{\uparrow,\downarrow\}$. Hopping
dimerization in one of the chains can be introduced by using a secondary
lattice on an already isolated ladder obtained using the primary
lasers, as sketched in Fig.~\ref{fig:scheme}~(a).
The Hubbard interaction can be simulated by aligning the dipoles in
individual chains at the so-called magic angle, such that in-leg interactions vanish, and 
only inter-leg interactions are relevant.

Alternatively, state dependent optical
lattices~\cite{soltan2011multi,yang2017spin} may allow for a
direct experimental realization of Model~\eqref{eq:hsshh}. The latter may be extended to more than two components, for which topological inheritance 
may occur as well. As an example, Fig.~\ref{fig:scheme}~(b) shows our results for a three-component system (assuming
SU(3)-symmetric all-to-all interactions, $U_{SU(3)}$), in which only one component is topological. As for the two-component case,   
the trivial components fully inherit the topological properties for relatively weak interactions.


\begin{acknowledgments}
The  computational simulations were carried out using the Param-Ishan HPC facility at Indian Institute of Technology - Guwahati, India. 
S.G. acknowledges support by the Swiss National Science Foundation under Division II. 
T.M. acknowledges DST-SERB, India for the early career grant through Project No.  ECR/2017/001069.
L. S. acknowledges support of the Deutsche Forschungsgemeinschaft (DFG, German Research Foundation) under Germany's Excellence Strategy -- EXC-2123 QuantumFrontiers -- 390837967.
\end{acknowledgments}


\bibliography{references}

\end{document}